\documentclass[journal]{IEEEtran}
\usepackage{float}
\usepackage{graphicx}
\usepackage{epsfig}
\usepackage{bm}
\usepackage{amsmath}
\usepackage{subfigure}
\usepackage{array}
\usepackage{booktabs}
\usepackage{CJK}
\usepackage{cases}
\usepackage{subeqnarray}
\usepackage{amsfonts}
\usepackage{amssymb}
\usepackage{diagbox}
\usepackage{algorithm}
\usepackage{algorithmic}
\usepackage{amsthm}
\usepackage{color}
\usepackage{ulem}
\usepackage{setspace}
\newcounter{MYtempeqncnt}

\title{\huge{Accumulate Then Transmit: Multi-user Scheduling in Full-Duplex Wireless-Powered IoT Systems}}

\author{\IEEEauthorblockN{Di Zhai, He Chen, Zihuai Lin, Yonghui Li and Branka Vucetic}
\thanks{The authors are with School of Electrical and Information Engineering, The University of Sydney, Sydney, NSW 2006, Australia (email: di.zhai@sydney.edu.au, he.chen@sydney.edu.au, zihuai.lin@sydney.edu.au, yonghui.li@sydney.edu.au, branka.vucetic@sydney.edu.au).}
}

\theoremstyle{plain}
\newtheorem{Prop1}{Proposition}
\begin{document}
\maketitle
\pagestyle{empty}  
\thispagestyle{empty} 
\begin{abstract}
This paper develops and evaluates an accumulate-then-transmit framework for multi-user scheduling in a full-duplex (FD) wireless-powered Internet-of-Things system, consisting of multiple energy harvesting (EH) IoT devices (IoDs) and one FD hybrid access point (HAP). All IoDs have no embedded energy supply and thus need to perform EH before transmitting their data to the HAP. Thanks to its FD capability, the HAP can simultaneously receive data uplink and broadcast energy-bearing signals downlink to charge IoDs. The instantaneous channel information is assumed unavailable throughout this paper. To maximize the system average throughput, we design a new throughput-oriented scheduling scheme, in which a single IoD with the maximum weighted residual energy is selected to transmit information to the HAP, while the other IoDs harvest and accumulate energy from the signals broadcast by the HAP. However, similar to most of the existing throughput-oriented schemes, the proposed throughout-oriented scheme also leads to unfair inter-user throughput because IoDs with better channel performance will be granted more transmission opportunities. To strike a balance between the system throughput and user fairness, we then propose a fairness-oriented scheduling scheme based on the normalized accumulated energy. To evaluate the system performance, we model the dynamic charging/discharging processes of each IoD as a finite-state Markov Chain. Analytical expressions of the system outage probability and average throughput are derived over Rician fading channels for both proposed schemes. Simulation results validate the performance analysis and demonstrate the performance superiority of both proposed schemes over the existing schemes.
\end{abstract}
\section{Introduction}
The Internet-of-Things (IoT), as an intelligent infrastructure, is expected to be integrated in the fifth-generation cellular systems to improve our daily life. To this end, massive and ubiquitous wireless sensors will be deployed [$\ref{IOT_EH1}$]. One of the main challenging problems to make the proliferation of IoT a reality is to supply adequate energy to maintain the system operation in a self-sufficient manner without compromising the system performance. Thereby, it is crucial to improve the operation lifetime of various sensors in IoT systems. Although numerous efforts, such as using lightweight routing protocols or equipping low-power radio transceivers, have been made to achieve this goal, wireless energy harvesting (EH) technique has recently been proposed as a new viable solution to prolong battery longevity [$\ref{IOT_EH2}$]-[$\ref{IOT_EH4}$].

With this technique, wireless devices in IoT systems can harvest energy from radio frequency (RF) signals emitted by ambient or dedicated transmitters and rely on the harvested energy to perform information transmission/processing.
However, RF-enabled EH has not been widely used in practical IoT systems due to the severe propagation attenuation of RF signal power. Fortunately, with the latest breakthroughs in wireless communications, namely small cells [$\ref{Small_cell}$], the application of large-scale antenna arrays (e.g., massive MIMO) [$\ref{massive_MIMO}$] and millimeter-wave communications [$\ref{mmWave}$], more attentions are paid in the field of short-distance communications, which make the RF-enabled EH more feasible than before. Furthermore, the energy consumption of IoT devices will continually reduce by the advanced circuit design, which makes the harvested energy sufficient to support their operation [$\ref{low_power}$]. Thus, it is believed that the RF-enable EH will be implemented widely in the near future [$\ref{IOT_EH2}$].

RF-enabled EH technique has opened a new research paradigm, termed wireless-powered communication network (WPCN) [$\ref{Introduction_WPCN}$]. In a WPCN, wireless devices are purely powered by a dedicated wireless energy transmitter (WET) in the downlink (DL) and transmit their information using the harvested energy in the uplink (UL). Instead of using the natural energy, the dedicated WET makes the EH process more controllable. The design of WPCNs for different setups has drawn tremendous interest recently [$\ref{WPCN-Rui}$]-[$\ref{WPCN-HTC}$]. In [$\ref{WPCN-Rui}$], a multi-user WPCN was investigated and a harvest-then-transmit (HTT) protocol was proposed. In the HTT protocol, within each transmission block, the users first harvest energy from the RF signals broadcast by a single-antenna hybrid access point (HAP) in the DL and then transmits information to the HAP in the UL in a time division multiple access (TDMA) manner. To maximize the system sum-throughput in each transmission block, the duration of both DL EH and UL information transmission (IT) was jointly optimized according to the channel quality of each link. Reference [$\ref{WPCN-Rui2}$] extended [$\ref{WPCN-Rui}$] to the scenario with a multi-antenna HAP. The objective of [$\ref{WPCN-Rui2}$] was to maximize the minimum throughput among all users through jointly optimizing DL-UL time allocation, DL energy beamforming and receiving beamforming. A power beacon-assisted WPCN (PB-WPCN) system was introduced in [$\ref{WPCN-PB}$], wherein besides the HAP, the power beacon can also help charging the users. In [$\ref{WPCN-PB}$], the authors designed a paid incentive mechanism based on game theory to encourage PBs to charge the users and the objective was to maximize the weighted sum-throughput of all HAP-user pairs.
In addition to the aforementioned setups, the wireless-powered node can also be used as a relay to enlarge the network coverage, forming the so-called wireless-powered cooperative communication network (WPCCN). Along this direction, the authors of [$\ref{WPCN-HTC}$] introduced a harvest-then-cooperative (HTC) protocol, in which the source node (SN) and relay first harvest energy from the HAP, then work cooperatively for the SN's IT in the UL. Moreover, due to the requirements of high reliability and stringent latency in emerging wireless applications, the short-packet WPCN has been studied in  [$\ref{URLLC1}$]-[$\ref{URLLC3}$] recently.

On the other hand, as one of the important techniques in the upcoming 5G cellular systems, full-duplex (FD) communications have received growing interests recently, see, e.g., [$\ref{FD_introduction}$], [$\ref{FD_WPCN_BF}$] and references therein. In a FD wireless system, devices can transmit and receive data simultaneously in the same frequency band while the receiving antenna will suffer from self-interference (SI). In the existing literature, there have been some initial efforts on the design of FD-WPCN. Authors of [$\ref{FD-WPCN-Rui}$] studied a time division (TD) based FD-WPCN, wherein a HAP first transfers RF energy to multiple users and receives data from each user via allocating a certain fraction of each transmission block to them. The weighted sum throughput of the considered FD-WPCN was maximized by optimizing the time allocation among EH and information transmission of all users. Similar to [$\ref{FD-WPCN-Rui}$], reference [$\ref{FD-WPCN-Sun}$] also maximized the system weighted sum throughput of a multi-user FD-WPCN by taking into account the energy causality constraint, for which the energy can be consumed only after it has been harvested. The SI in [$\ref{FD-WPCN-Sun}$] was assumed to be removed perfectly. With the same system setup, the impact of imperfect SI cancelation was investigated in [$\ref{FD-WPCN-Sun2}$]. It is also worth mentioning that the antenna pair selection and two-way information flow issues of single-user FD-WPCNs were respectively investigated in [$\ref{FD-WPCN-Mingjin}$], [$\ref{FD-WPCN-Zihao}$]. Moreover, wireless-powered FD relay networks were studied in [$\ref{FD-WPCN-Relay1}$]-[$\ref{FD-WPCN-Relay3}$] and a wireless-powered FD friendly jamming protocol was designed and analyzed in [$\ref{FD-WPCN-Jamming}$].

\subsection{Motivation}
For a multi-user FD-WPCN, the aforementioned TD-based scheduling may lead to a sub-optimal system performance as the amount of harvested energy is normally very limited during each scheduled slot and could not support an effective IT. In this case, the EH devices may harvest and accumulate energy for several consecutive blocks before being scheduled to perform one IT. This is in contrast to the TD-based scheduling schemes in [$\ref{FD-WPCN-Rui}$], [$\ref{FD-WPCN-Sun}$], where the SNs exhaust the harvested energy to perform instantaneous IT during the scheduled time slot of each transmission block. Hence, the inherent energy accumulation (EA) process introduced by single user selection in FD-WPCNs should be carefully incorporated into the system design and evaluation. For example, the harvested energy in the current transmission block could be saved for the future usage if the SN cannot support an effective IT. This nature makes the existing best node selection schemes designed for conventional systems no longer applicable. Note that the EA processes of a wireless-powered multiple input single output (MISO) system with single user and a wireless-powered relay network with multi-relay selection were characterized in [$\ref{EA1}$], [$\ref{EA2}$]. However, to our best knowledge, the multi-user scheduling in FD-WPCNs with the inherent EA processes has not been studied in existing literature.

Motivated by this gap, in this paper we study a FD wireless-powered IoT (WP-IoT) system consisting of one FD HAP and multiple EH IoT devices (IoDs). The HAP has constant power supply and keeps broadcasting RF signals to charge IoDs wirelessly, while the IoDs are wireless-powered devices such that they accumulate the harvested energy from RF signals broadcast by the HAP in DL before being scheduled to transmit their data to the HAP in UL. Now a natural question arises in the considered WP-IoT system, that is, `\textit{Which IoD should be selected to transmit information in the UL considering that the accumulated energy for each individual IoD at different locations could be quite distinct?}'

In conventional multi-user scheduling schemes, to maximize the system average throughout, the schedulers intend to select the user with the best instantaneous channel gain [$\ref{bestH}$]. However, as described in [$\ref{PF1}$]-[$\ref{PF4}$], the scheduling policy aiming to maximize the system average throughput under multi-user scenario can be unfair, as the wireless devices with the poor channel quality may be starved. To improve the system fairness of the conventional multi-user networks, the authors of [$\ref{PF1}$] proposed to select the user which has the largest ratio between the instantaneous received SNR and average received SNR. In [$\ref{hybridPF}$], a hybrid multi-user scheduling scheme to balance the system throughput and fairness was developed. The works of [$\ref{PF2}$] and [$\ref{PF4}$] improved the system fairness for a WPCN by setting higher weight factor for the devices with poor channel quality. However, due to extreme high complexity, the policies in [$\ref{PF2}$] and [$\ref{PF4}$] are hard to implement in an online manner. Moreover, as explained in [$\ref{Channel_training1}$] and [$\ref{Channel_training2}$], realizing channel estimation in a WPCN system is not an easy task. This is because the wireless-powered devices could be low-cost and low-complexity nodes such that they may have no capability to conduct accurate channel estimation, which further disenables the policies mentioned above.

To address this issue, in this paper we develop a new accumulate-then-transmit framework for the considered WP-IoT system by proposing two new user scheduling policies focusing respectively on average throughput and user fairness under a practical assumption that the CSIT (channel state information at the transmitter) is unknown. Motivated by this assumption, we propose to use the instantaneous energy state information at each IoD and the statistical channel knowledge of each link to schedule multiple IoDs.

To model the EA process and the dynamic charging/discharging behavior, a discritized-state battery model is required. We follow [$\ref{EA1}$], [$\ref{EA2}$] and [$\ref{iid1}$], and adopt the finite-state Markov chain to model the EA process and characterize the battery steady states. Furthermore, different from those existing work wherein the steady state of each node can be determined independently, in our model, the steady state of one IoD will be affected by that of all other IoDs for competing for the limited UL spectrum resources, which makes the separated performance analysis of each IoD no longer applicable. This calls for a new framework to evaluate the performance of the proposed schemes.
 \subsection{Our Contributions}
 The main contributions of this paper are summarized as follows:
\begin{itemize}
   \item We develop a new throughput-oriented multi-user scheduling scheme for the considered FD WP-IoT system with imperfect SI cancellation. In our scheme, at the beginning of each transmission block, the IoD with the maximum weighted residual energy is selected to transmit its information to the HAP in UL, while all other IoDs perform EH operation and accumulate the harvested energy for future IT.
   \item To cater for IoDs with poor channel condition, we then propose a fairness-oriented multi-user scheduling scheme for the FD WP-IoT system. In this scheme, at the beginning of each transmission block, the scheduler will select the IoD with the largest ratio between the exact accumulated energy and the average accumulated energy within the time period from the latest data transmission to the current block.
   \item By considering that all IoD-HAP links experience practical \textit{independent but non-identical distribution} channel fading, we analyze the system average throughput for both proposed scheduling policies by modeling the EA processes of all IoDs as finite-state Markov chains (MCs). It is worth pointing out that IoD selection is carried out by jointly considering all IoDs' energy states. The state transition matrices of all IoDs' MCs are thus tangled together, which makes the analytical performance analysis of the considered system nontrivial. The steady state distribution of IoDs' battery is shown to be the root of a complex multi-variable equation set and can be solved through the fixed-point iteration method. All theoretical analysis is validated by Monte Carlo simulations. Numerical results show that the fairness-oriented policy can provide better throughput performance than the existing round robin policy while holds similar fairness. On the other hand, by sacrificing the system fairness, the throughput-oriented policy can achieves higher throughput than the fairness-oriented policy.
 \end{itemize}
\begin{figure}[t]
\includegraphics[height=2.0in]{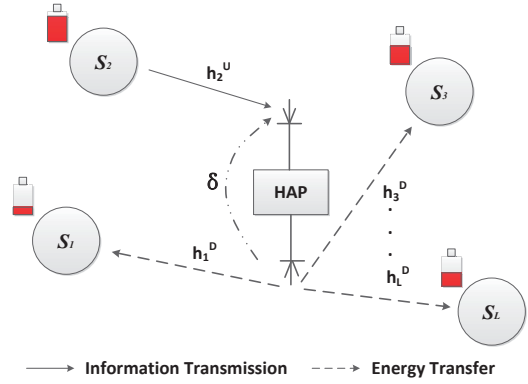}
\centering
\caption{System model of the considered FD WP-IoT system with one HAP and $L$ wireless-powered IoDs. Suppose $S_2$ is the selected node for data transmission in the UL, and the rest $L-1$ devices harvest energy in the DL.}
\end{figure}

 A complete list of acronyms can be found in Table I and the rest of the paper is organized as follows: Section II presents the system model and channel model for the considered FD WP-IoT system. In Section III, we elaborate the scheme design for the throughput-oriented policy in detail in the FD WP-IoT system. The closed-form expressions of system outage probability and average throughput for FD WP-IoT system are derived. In Section IV, another fairness-oriented policy is introduced in detail. In Section V, numerical results are presented to validate the analysis. Finally, Section VI concludes the paper.
 \begin{table}[t]
  \begin{center}
    \caption{List of Acronyms}
    \label{tab:table1}
    \begin{tabular}{ll} 
      \textbf{Symbol} & \textbf{Description} \\
      \hline
      ARQ & Automatic Repeat Request \\
      AWGN & Additive white Gaussian noise \\
      CDF & Cumulative Distribution Function \\
      CSI & Channel State Information \\
      CSIT & Channel State Information at Transmitter\\
      DL & DownLink \\
      EA & Energy Accumulation \\
      EH & Energy Harvesting \\
      FD & Full-Duplex \\
      FI & Fixed-point Iteration \\
      HAP & Hybrid Access Point \\
      HD & Half-Duplex \\
      HTC & Harvest-Then-Cooperative \\
      HTT & Harvest-Then-Transmit \\
      IoD & Internet-of-things Device \\
      IoT & Internet-of-Things \\
      IT & Information Transmission \\
      MC & Markov Chain \\
      PB & Power Beacon \\
      RF & Radio Frequency \\
      RR & Round Robin \\
      RS & Random Selection \\
      SI & Self-Interference \\
      SINR & Signal-to-Interference-plus-noise-Ratio \\
      SN & Source Node \\
      UL & UpLink \\
      WPCN & Wireless-Powered Communication Network\\
      WP-IoT & Wireless-Powered-Internet-of-Things \\
      \hline
    \end{tabular}
  \end{center}
\end{table}
\section{System Model}
\subsection{System Model}
As shown in Fig. 1, we consider a multi-user FD WP-IoT system consisting of one FD HAP and multiple half-duplex (HD) IoDs. We assume that all IoDs are equipped with one antenna, while the HAP is equipped with two antennas. The HAP's two antennas structure enables its FD working mode. Particularly, one antenna is used to receive the scheduled IoD's signal in the UL and the other one is used to wirelessly charge the remaining IoDs by broadcasting RF signals in the DL. Moreover, the HAP is assumed to be connected to an external energy supply (e.g., the power grid), while all the IoDs are wireless-powered devices and purely rely on the energy harvested from RF signals broadcast by the HAP to support their operation. Besides, all IoDs are equipped with separated energy and information receivers and share the same antenna, which indicates that each IoD can only work in either EH mode or IT mode. As such, they can flexibly switch between an EH mode and an IT mode at the beginning of each transmission block according to the scheduling policy. We also assume that each IoD is equipped with a finite-capacity rechargeable battery such that it can perform EA and schedule the accumulated energy across various transmission blocks.

Multi-user scheduling is applied in the considered FD WP-IoT system. Specifically, within each transmission block, at most one IoD is chosen to operate in the IT mode, while the IoDs fail to be chosen are in the EH mode. At each IoD operating in the EH mode, the received signal is passed to the energy receiver to convert to direct current and charge the battery. In contrast, if a certain IoD is selected to operate in the IT mode, it will consume its accumulated energy to transmit its information to the HAP in the UL. Note that because both UL and DL work at the same band and the HAP operates in a FD mode, the received signal from the IoD operating in the IT mode at the HAP will suffer from the SI caused by broadcasting of energy-bearing signals in the DL.

We hereafter use $S_i$, $i \in \{1,2,\ldots,L\}$, to denote the $i$-th IoD and $L$ is the total number of IoDs. To the authors' best knowledge, the up-to-date wireless energy transfer techniques could only be operated within a relatively short communication range such that the line-of-sight (LoS) path is very likely to exist in these links. In this sense, the Rician fading would be the most appropriate model to characterize the channel fading of all links in the considered FD WP-IoT system. We thus consider that the channel coefficients of the UL and DL links between $S_i$ and HAP, denoted by $h^U_i$ and $h^D_i$ respectively, follow independent and identical Rician distribution with the Rician factor $\Psi$, which is defined as the ratio of the powers of the LoS component to the scattered components, and average channel power gain $\bar{H_i}$. Besides, all channels in the system are assumed to experience frequency-flat and slow fading such that the instantaneous channel gains remain unchanged within each transmission block but change independently from one block to the other. Without loss of generality, we use $T$ to denote one transmission block with $T = 1$ hereafter.
\subsection{Energy Harvesting Phase}
Due to the HD constraint of IoDs, they can only operate in either EH or IT mode. When the $i$-th IoD works in the EH mode, it will harvest energy from the RF signals broadcast by the HAP during the entire transmission block and accumulate the harvested energy in the battery for future usage. The amount of harvested energy for a certain EH block at $S_i$ can be expressed as
\begin{equation}
\tilde{E_i}=\eta P_H H^D_iT,
\end{equation}
where $0<\eta<1$ is the energy conversion efficiency and $H^D_i = |h^D_i|^2$ is the DL instantaneous channel power gain of $S_i$. $P_H$ is the HAP transmit power. Note that in (1), we ignore the amount of energy harvested from the noise and the transmitted signal emitted by the IoD in the IT mode. The reason is that for the energy limited nature of the EH IoDs, the transmit power of the EH IoD is normally small, which is negligible compared to the transmit power of the HAP. In this sense, the obtained system performance can serve as a lower bound of practical systems. Note that the IoD operating in IT mode cannot harvest energy due to its HD constraint.
\subsection{Information Transmission Phase}
For the case that $S_i$ is selected to transmit information in the UL, let $P_i$  and $x_i$ denote its transmit power and transmitted symbol with $\mathbb{E}[|x_i|^2]=1$ respectively, where $\mathbb{E}[\cdot]$ represents the expectation operator. When the HAP receives the signal from $S_i$, its receiving antenna will overhear the energy-bearing signal broadcast by the transmitting antenna, which will cause the SI. Thus, the received signal at the HAP is a combination of the signal transmitted by $S_i$, SI and receiver noise, which can be expressed as
\begin{equation}
y_H = \sqrt{P_i}h^U_ix_i + \sqrt{P_H}\delta c + n,
\end{equation}
where $\delta$ is the residual SI that remains after imperfect SI elimination [$\ref{FD2}$], and $c$ is the energy-bearing symbol broadcast by the HAP in the DL with $\mathbb{E}[|c|^2]=1$. $c$ is designated only for energy transmission and thus can be chosen to be deterministic. $n$ denotes the additive white Gaussian noise (AWGN) with zero mean and variance $N_0$ at the HAP. The signal-to-interference-plus-noise ratio (SINR) at the HAP of the $i$-th IoD is thus given by
\begin{equation}
\label{SINR}
\gamma_i = \frac{P_iH^U_i}{N_0 + P_H \alpha},
\end{equation}
where the loop interference power gain $\alpha = |\delta|^2$ and the UL instantaneous channel power gain $H^U_i=|h^U_i|^2$ of $S_i$. In this paper, $\alpha$ is assumed to be constant. This is motivated by the fact that the energy-bearing signal can be effectively reduced by implementing isolation between the two HAP antennas or advanced analog and digital SI cancellation, such that the SI can be reduced to a negligible level or noise level, and the randomness can be suppressed dramatically [$\ref{FD2}$]-[$\ref{FD5}$].

Since multiple IoDs share the limited spectrum resource in the considered multi-user system, it is natural to ask ``\textit{which IoD should be selected to perform IT in a certain transmission block?}'' This is actually a non-trivial question to answer. This is because in order to achieve certain system object (i.e., maximize the system average throughput [$\ref{WPCN-Rui}$], [$\ref{FD_WPCN_BF}$] or maximize system fairness [$\ref{WPCN-Rui2}$], [$\ref{PF1}$]-[$\ref{PF4}$]), the designer needs to schedule different IoDs which have different capacities (i.e., different amount of residual energy, distinct transmission rate requirements and various channel quality). Hence, before selecting a proper IoD at the beginning of each transmission block, it is necessary to define a specific target for the scheduler of the system. In this context, we will introduce two user scheduling policies for both the throughput-oriented and fairness-oriented scenarios in the proposed FD WP-IoT system in the subsequent two sections.

%
%

\section{Design and Analysis of the Throughput-oriented Policy}
In this section, we first explain the principle of the proposed throughput-oriented scheduling policy. Then, we will elaborate how to use the Markov Chain to model the discretized IoD battery states. Finally, we mathematically describe the IoD behaviors and evaluate the system performance.
\subsection{Policy Design}
   To design the scheduling policy in the considered FD WP-IoT system, the widely used instantaneous CSI-based user scheduling schemes are no longer applicable due to the assumption of unknown instantaneous CSI [$\ref{bestH}$]-[$\ref{PF4}$]. On the other hand, considering the inherent EA process, the residual energy accumulated in the battery can be adopted as a feasible criterion to measure the capability of information transmission of each IoD. Motivated by this, we develop a weighted residual energy-based multi-user scheduling scheme to maximize the system average throughput for the scenarios without the instantaneous CSI. Besides, we assume at most one IoD is allowed to transmit data in the UL within one transmission block. To elaborate the policy of IoD selection, we first define the average throughput by the $i$-th IoD if it is selected to transmit information, which is given by
   \begin{equation}
    \bar{\Omega}_i = R(1-\bar{P}_{i,out})T,
   \end{equation}
where $i$ is the index of the IoD. $R$ is the rate requirement\footnote{For simplicity, we consider that the IoDs within the system have the same rate requirement. Our framework can also be extended to the case with distinct rate requirement.} and $\bar{P}_{i,out}$ is the data transmission outage probability of the $i$-th IoD. In this sense, to maximize the system average throughput, we need to select the IoD which possesses the lowest transmission outage probability. Recall that the channels between various IoDs and HAP are assumed to undergo independent and non-identical Rician fading. When the $i$-th IoD is selected to perform IT, $\bar{P}_{out}$ can be expressed as 
\begin{small}
\begin{equation}\label{eq:example_out}
\begin{split}
\bar{P}_{out} &= \textrm{Pr}\bigg(\log_2\Big(\frac{P_iH^U_i}{N_0 + P_H \alpha}+1\Big)<R\bigg)\\
& =  \textrm{Pr}\Big(H^U_i<\frac{(N_0 + P_H \alpha)(2^{R}-1)}{P_i}\Big)\\
& = F_{H^U_i}\Big(\frac{(N_0 + P_H \alpha)(2^{R}-1)}{P_i}\Big)\\
& = 1 - \mathrm{Q}_1\bigg(\sqrt{2\Psi},\sqrt{\frac{2(N_0 + P_H \alpha)(2^{R}-1)(\Psi+1)}{P_i\bar{H}_i}}\bigg),
\end{split}
\end{equation}
\end{small}
where $F_{H^U_i}(\cdot)$ is the CDF of $H^U_i$, which can be expressed as
\begin{equation}\label{eq:UL-DL-CDF}
F_{H^U_i}(x) =  1-\mathrm{Q}_1\left(\sqrt{2\Psi},\sqrt{\frac{2(\Psi+1)x}{\bar{H_i}}}\right),
\end{equation}
where $\mathrm{Q}_1(\cdot,\cdot)$ is the generalized first-order Marcum $\mathrm{Q}$-function [$\ref{math_tool}$] and we can verify that $\mathrm{Q}_1(Y_1,Y_2)$ is a monotonically decreasing function of $Y_2$. As such, the lowest $\bar{P}_{out}$ can be achieved statistically by selecting the IoD with the maximum $P_i\bar{H}_i$. On the other hand, due to the unknown instantaneous CSI, the selected IoD $S_i$ ought to exhaust its residual energy to transmit data to eliminate outage, i.e., $P_i = \frac{r_i[t]}{T}$ with $r_i[t]$ denoting $S_i$'s residual energy at the beginning of $t$-th transmission block. Overall, in our policy, the $i$-th IoD will be selected to transmit data during the $t$-th block if $i$ meets
\begin{equation}\label{condition}
i = \arg \max_{j\in \{1,2,\ldots,L\}}\Big\{{r_j}[t]\bar{H}_j\Big\}~\textrm{and}~{r_i}[t] \neq 0.
\end{equation}Under this assumption, if the current information transmission fails, the automatic repeat request (ARQ) procedure cannot be performed until the IoD accumulates enough energy and is scheduled to transmit information again. Considering the timeliness of the information, it is reasonable to declare a package loss instead of using ARQ in the considered WP-IoT system. Besides, we assume the transmit power of these IoDs can rarely exceed the working range of the power-related chips since the harvested energy are normally limited.
It is worth pointing out that each IoD only knows its own weighted residual energy, and is unaware of that of other IoDs, requiring the proposed scheduling policy to be implemented in a distributed way. This can be achieved through the time-backoff scheme. Specifically, all IoDs are synchronized by the HAP and at the beginning of each transmission block, each IoD sets a timer independently according to its own weighted residual energy. The timer of the $i$-th IoD at the beginning of the $t$-th time block is set to be inverse proportional to its weighted residual energy $r_i[t]\bar{H}_i$. The timer of the IoD with the maximum weighted residual energy will expire firstly and this selected IoD will broadcast a short flag packet to signal its presence. After hearing the first flag packet from a certain IoD, the HAP will broadcast it to all the remaining IoDs to declare the selection of this IoD in the current round. All other IoDs will switch to the energy harvesting mode to harvest energy from the HAP. In this case, although the synchronization and final selection decision are done by the HAP, the entire user scheduling process is mostly conducted at the IoD side in a distributed way. As such, the complexity and computational use of resources will almost not scale as the number of IoDs increases.

For simplicity, we neglect the time and energy consumed by the user scheduling process based on the following considerations. The advanced performance of the MCU used in IoT systems makes the time consumption of the timer setting negligible. On the other hand, the state-of-the-art ultra-low power technology of the MCU makes the energy consumption of the circuit negligible compared to that used for information transmission.
\subsection{Markov Chain for IoD's Batteries}
To characterize the performance of the proposed scheduling policy, we adopt a discrete-level and finite-capacity battery model [$\ref{EA1}$]. It is thus natural to use a finite-state MC to model the dynamic charging/discharging behaviors of IoD's batteries. Note that in the proposed scheduling policy, the user selection procedure depends on the energy status of all IoDs. Thus, the state transitions and their associated steady state distributions of all IoD batteries are correlated with each other and thus cannot be evaluated separately, which makes the theoretical analysis non-trivial.

 Denote by $C$ the capacity of each IoD battery and by $K$ the number of discrete energy levels excluding the empty level in each battery\footnote{Note that the proposed analysis framework can be extended to the case with distinct capacity and energy levels.}. Then, the $k$-th energy level of each IoD's battery can be presented as $\varepsilon_k = \frac{kC}{K}, k \in \{0,1,2,\ldots,K\}$. It is worth pointing out that the adopted discrete battery model can tightly approximate its continuous counterpart when the number of energy levels (i.e., $K$) is large enough, as shown in [$\ref{iid1}$]. The transition probability $T_i^{k,l}$ is defined as the probability of the transition from state $k$ to state $l$ at the $i$-th IoD. With the adopted discrete-level battery model, the amount of harvested energy and the residual energy can only be one of the discrete energy levels. Thus, the discretized amount of harvested energy at the $i$-th IoD during one EH operation is defined as
\begin{equation}
\label{discret}
E_i \triangleq \varepsilon_l,~\textrm{where}~l = \arg \max_{k\in\{0,1,\ldots,K\}}\Big\{\varepsilon_k:\varepsilon_k \leq \tilde{E_i}\Big\}.
\end{equation}

We are now ready to describe the IoD behaviors in the proposed throughput-oriented policy mathematically. Let $\zeta_i[t] \in \{\zeta_T, \zeta_E\}, t = 1,2,3,\ldots$, denote the operation mode of the $i$-th IoD during the $t$-th transmission block, where $\zeta_T$ and $\zeta_E$ denote the IT and EH modes respectively. According to the throughput-oriented policy, if the $i$-th IoD is scheduled to operate in the IT mode, $S_i$ should meet the condition ($\ref{condition}$). We thus have
\begin{equation}
\label{FD_states}
\zeta_i[t]=
\begin{cases}
\zeta_T,~\textrm{if}~(\ref{condition})~\textrm{holds}\\
\zeta_E,~\textrm{otherwise}\\
\end{cases}
.
\end{equation}
Moreover, we define $\Phi_i[t]=r_i[t]\bar{H}_i$ as the weighted residual energy of $S_i$ for notation simplicity, and $\Phi_i[t]$ evolves to $\Phi_i[t+1]$ as follows
\begin{equation}
\label{E_changed}
\Phi_i[t+1]=
\begin{cases}
0,~\textrm{if}~\zeta_i[t]=\zeta_T\\
\bar{H}_i\min \Big\{r_i[t]+E_i,C \Big\}, \textrm{if}~\zeta_i[t]=\zeta_E\\
\end{cases}
.
\end{equation}
Noted that in ($\ref{E_changed}$), we consider that the energy consumption at the IoDs is dominated by their IT operation, and other types of energy consumption (e.g., signal processing) is assumed to be negligible for simplicity.

\begin{figure}[t]
\includegraphics[height=3.5in]{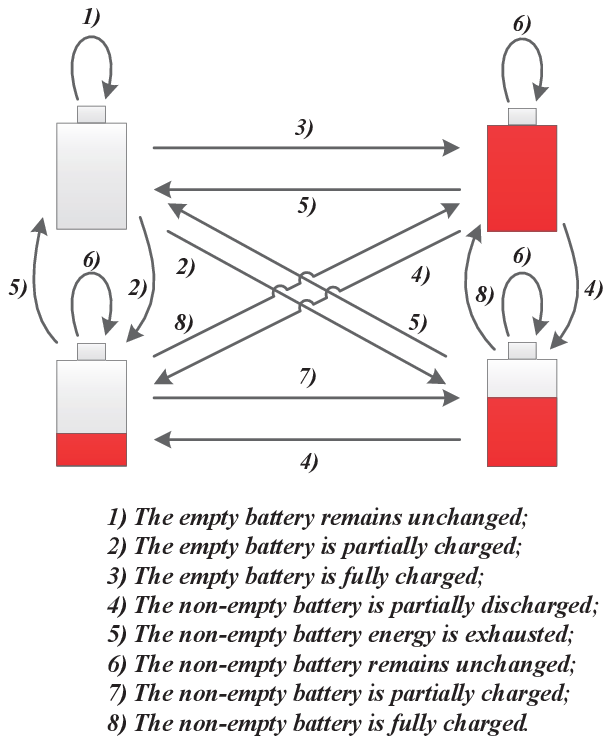}
\centering
\caption{\label{state_transition_TO}The different cases of state transition in throughput-oriented policy.}
\end{figure}

With the mathematical description of the proposed scheduling policy and the MC model defined above, we can proceed to evaluate the state transition probabilities of the MC for each IoD. As Fig. $\ref{state_transition_TO}$ shows, the transition probabilities can be summarized into \textit{eight} different cases depending on different initial state $k$ and end state $l$: 1) The empty battery remains unchanged; 2) The empty battery is partially charged; 3) The empty battery is fully charged; 4) The non-empty battery is partially discharged; 5) The non-empty battery energy is exhausted; 6) The non-empty battery remains unchanged; 7) The non-empty battery is partially charged; 8) The non-empty battery is fully charged. Mathematically, the transition probability of $i$-th IoD from state $k$ to state $l$ can be expressed as
\begin{small}
\begin{equation}\label{FD_T}
T^{k,l}_i=
\begin{cases}
\textrm{Pr}(\tilde{E_i} \leq \frac{C}{K}),~\textrm{if}~k = l=0\\
\textrm{Pr}\Big(\frac{(l-k)C}{K}\leq \tilde{E_i} \leq \frac{(l-k+1)C}{K}\Big),\\
~~~~~~~~~~~~~~~~~~\textrm{if}~k < l < K,~k=0\\
   \textrm{Pr}\Big(\tilde{E_i} \geq \frac{(K-k)C}{K}\Big),~\textrm{if}~k=0,~l = K\\
   0,~\textrm{if}~k > l,~l>0\\
   \upsilon^{\mathrm{T}}_{i,k},~\textrm{if}~k>0,~l=0\\
   (1-\upsilon^{\mathrm{T}}_{i,k})\textrm{Pr}(\tilde{E_i} \leq \frac{C}{K}),~\textrm{if}~k >0,~k = l \neq K\\
   (1-\upsilon^{\mathrm{T}}_{i,k})\textrm{Pr}\Big(\frac{(l-k)C}{K}\leq \tilde{E_i} \leq \frac{(l-k+1)C}{K}\Big),\\
   ~~~~~~~~~~~~~~~~~~~\textrm{if}~0<k < l \neq K\\
   (1-\upsilon^{\mathrm{T}}_{i,k})\textrm{Pr}(\tilde{E_i} \geq \frac{(K-k)C}{K}),~\textrm{if}~k>0,~l = K
\end{cases}
,
\end{equation}
\end{small}where $\upsilon^{\mathrm{T}}_{i,k}$ represents the probability that the $i$-th IoD is scheduled to operate in the IT mode at battery level $k$ on throughput-oriented scheme.

We now explain how to calculate the probability terms in ($\ref{FD_T}$). We use the case $\textrm{Pr}(\tilde{E_i}\leq \frac{C}{K})$ as an example. From the definition of discretization given in ($\ref{discret}$), condition $\tilde{E_i} = \eta P_H H^D_i < \varepsilon_1 = C/K$ must hold such that the increment of harvested energy can be discretized to zero ($k=l=0$). With the help of CDF defined in ($\ref{eq:UL-DL-CDF}$), the probability of an effective zero harvested energy is given by
\begin{equation}
\label{cdf+1}
\textrm{Pr}\left(\tilde{E_i} \leq \frac{C}{K}\right) = F_{H^D_i}\left(\frac{C}{\eta P_H K}\right),
\end{equation}where $F_{H^D_i}(\cdot)$ is the CDF of $H^D_i$ and has the same expression with $F_{H^U_i}(\cdot)$.
The probability of other transition probability in ($\ref{FD_T}$) can be evaluated similarly, which are omitted for brevity. In the subsequent subsections, we will first figure out the relationship between $\upsilon^{\mathrm{T}}_{i,k}$ and the steady state distribution of all IoDs, and then calculate the steady state distributions of all IoDs by applying the fixed-point iterative method [$\ref{Fixed_point}$].
\subsection{Steady State Calculation}
We use
\begin{equation*}
\bm{\pi}_i =[{\pi}_{i,0},\cdots,{\pi}_{i,k},\cdots,{\pi}_{i,K}]^T
\end{equation*}
to denote the battery steady state distribution of the $i$-th IoD. Specifically, $\pi_{i,k}$ denotes the probability that the battery state of $S_i$ is $k$. To calculate these steady states, we need the fully observable transition probability $T^{k,l}_i$ between different states. Thus, we first derive the relationship between $\upsilon^{\mathrm{T}}_{i,k}$ and stationary states of all IoDs' battery.

Recall that the event $S_i$ operating in the IT mode at the battery level $k$ happens only when the condition ($\ref{condition}$) is satisfied with $r_i[t]=\varepsilon_k$. Thus, $\upsilon^{\mathrm{T}}_{i,k}$ can be expressed as
\begin{equation}
\label{v_WRE_ik}
\upsilon^{\mathrm{T}}_{i,k} =\prod\limits_{p\in\varphi\setminus\{i\}}\Big(\sum\limits_{q\in\mathcal{G}^{\mathrm{T}}_{p,k}}\pi_{p,q}\Big),
\end{equation}
where $\mathcal{G}^{\mathrm{T}}_{p,k}$ is the set including all battery levels of the $p$-th IoD that are smaller than $\frac{\varepsilon_{k}\bar{H}_i}{\bar{H}_p}$ and $\varphi = \{1,2,\ldots,L\}$ is the set of all IoDs. Here we use $\varphi\setminus \{i\}$ to denote the relative complement of $i$ with respect to the set $\varphi$, i.e., $\varphi\verb|\|\{i\}=\{1,2,\ldots,i-1,i+1,\ldots,L\}$.

We are ready to calculate the battery steady state distribution of different IoDs after we have ($\ref{v_WRE_ik}$). We use $\mathbf{Z}_{i}$ to denote the battery state transition matrix of the $i$-th IoD. We can verify that the MC for each IoD's battery has a unique steady state (ergodic between different states and no absorbing state exists), which should satisfy the following equation [$\ref{markov_chain}$]
\begin{equation}
\label{FP1}
\bm{\pi}_i = \mathbf{Z}^T_i\bm{\pi}_i.
\end{equation}The steady state distribution of $S_i$'s battery can be solved from ($\ref{FP1}$) as [$\ref{iid1}$]
\begin{equation}\label{solution_MC}
\bm{\pi}_i = \Big(\mathbf{Z}^T_i-\mathbf{I}+\mathbf{B}\Big)^{-1}\mathbf{b},
\end{equation}where $\mathbf{B}_{i,j} = 1, \forall i,j$ and $\mathbf{b} = (1,1,\cdots,1)^T$. However, through
($\ref{v_WRE_ik}$), we find the state transition matrix of certain IoD is related with that of other IoDs by noting that $\upsilon^{\mathrm{T}}_{i,k}$ is included in $\mathbf{Z}_{i}$. This indicates that the state transition matrices of all IoDs are inherently tangled together. As such, we cannot treat them separately and solve the individual steady state equations to calculate their steady state distributions. Fortunately, after a careful observation on ($\ref{FP1}$), we find that the battery steady state distribution of various IoDs constructs a multi-variable, high-order and non-linear equation set. The root of the equation set is the battery steady state of all IoDs. On the other hand, the fixed-point iteration (FI) approach has been widely applied to solve multi-variable, high-order and non-linear equation set [$\ref{Fixed_point}$]. Motivated by this, we introduce the FI method. To construct the iteration function, we first set
\begin{equation*}
\bm{\pi} = [\bm{\pi}_1^T,\ldots,\bm{\pi}_L^T]^T,
\end{equation*}
and
\begin{align*}
\mathbf{Z}^T=
\begin{bmatrix}
\begin{smallmatrix}
\mathbf{Z}^T_1 &  & & &\\
     & \ddots & & &\\
     &  &\mathbf{Z}^T_i & &\\
     &  & &\ddots &\\
    & & & &\mathbf{Z}^T_L \\
\end{smallmatrix}
\end{bmatrix}.
\end{align*}
Next we integrate all the IoDs with the same form as ($\ref{FP1}$), which is
\begin{equation}
\label{FP2}
\bm{\pi} = \digamma(\bm{\pi}),
\end{equation}
where
\begin{equation*}
\digamma(\bm{\pi}) = \mathbf{Z}^T\bm{\pi}.
\end{equation*}
Then, the iteration process can be written as [$\ref{Fixed_point}$]
\begin{equation}
\label{iteration}
\bm{\pi}^{(s+1)} = \digamma(\bm{\pi}^{(s)}),
\end{equation}
and $s$ denotes the iterative index.
\begin{Prop1}
 There exists a unique fixed point $\bm{\pi}^*$ and the iteration process in ($\ref{iteration}$) can converge to $\bm{\pi}^*$, if there exists a real number $\lambda$ ($0<\lambda<1$), which makes the inequation\footnote{The norm can be any norm. Here we consider 1-norm.}
\begin{equation}
\label{convergence_condition}
\| \digamma(\bm{\pi}) - \digamma(\bm{\pi}')\|_1 \leq \lambda\| \bm{\pi} - \bm{\pi}'\|_1, \forall ~\bm{\pi},\bm{\pi}' \in D
\end{equation}
always true. Here $D \subseteq [0,1]^{(K+1)\times L}$ is the domain of $\bm{\pi}$ and $\bm{\pi}'$. $\|\cdot\|_1$ denotes the 1-norm operation.
\end{Prop1}
\begin{proof}
Firstly we define $\bm{\pi} = [{\bm{\pi}_1}^T,\ldots,{\bm{\pi}_i}^T,\ldots,{\bm{\pi}_L}^T]^T$ and $\bm{\pi}' = [{\bm{\pi}'_1}^T,\ldots,{\bm{\pi}'_i}^T,\ldots{\bm{\pi}'_L}^T]^T$, where $\bm{\pi}_i = [\pi_{i,0},\ldots,\pi_{i,K}]^T$ and $\bm{\pi}'_i= [\pi'_{i,0},\ldots,\pi'_{i,K}]^T$. The elements in $\bm{\pi}_i$ and $\bm{\pi}'_i$ are all non-negative and $\|\bm{\pi}_i\| = \|\bm{\pi}'_i\| = 1$. Expanding the left side of inequality ($\ref{convergence_condition}$), we have
\begin{equation}
\label{CC2}
 \|\digamma(\bm{\pi}) - \digamma(\bm{\pi}')\|_1 = \sum\limits_{i=1}^L\sum\limits_{l=0}^{K}|\sum\limits_{k=0}^{K}T_i^{k,l}(\pi_{i,k}-\pi'_{i,k})|
\end{equation}For $S_i$ at end energy state $l$, it meets
\begin{equation}\label{CC3}
|\sum\limits_{k=0}^{K}T_i^{k,l}(\pi_{i,k}-\pi'_{i,k})|\leq \sum\limits_{k=0}^{K}T_i^{k,l}|(\pi_{i,k}-\pi'_{i,k})|.
\end{equation}
The equality may hold only if the end state $l \neq 0$ and $l \neq K$ since $\|\bm{\pi}_i\| = \|\bm{\pi}'_i\| = 1$. Hence, after doing the sum calculation for ($\ref{CC3}$), we can have
\begin{equation}\label{CC5}
\sum\limits_{l=0}^{K}|\sum\limits_{k=0}^{K}T_i^{k,l}(\pi_{i,k}-\pi'_{i,k})| < \sum\limits_{l=0}^{K}\sum\limits_{k=0}^{K}T_i^{k,l}|\pi_{i,k}-\pi'_{i,k}|.
\end{equation}On the other hand, expanding $\|\bm{\pi}_i - \bm{\pi}'_i\|_1$, we can have
\begin{equation}
\begin{split}
\|\bm{\pi}_i - \bm{\pi}'_i\|_1 &= \sum\limits_{k=0}^{K}|\pi_{i,k}-\pi'_{i,k}|\\ &\overset{(a)}{=}\sum\limits_{l=0}^{K}\sum\limits_{k=0}^{K}T_i^{k,l}|\pi_{i,k}-\pi'_{i,k}|,
\end{split}
\end{equation}where (a) follows $\sum\limits_{l=0}^{K}T_i^{k,l}=1$.
Thus, with ($\ref{CC5}$), we can conclude that for any $S_i$, there always exists a real number $\lambda_i$ ($0<\lambda_i<1$), which makes the inequation
\begin{equation}
\sum\limits_{l=0}^{K}|\sum\limits_{k=0}^{K}T_i^{k,l}(\pi_{i,k}-\pi'_{i,k})| \leq \lambda_i \|\bm{\pi}_i - \bm{\pi}'_i\|_1
\end{equation}true. We then further have
\begin{equation}\label{CC4}
\begin{split}
 \|\digamma(\bm{\pi}) - \digamma(\bm{\pi}')\|_1 &= \sum\limits_{i=1}^L\sum\limits_{l=0}^{K}|\sum\limits_{k=0}^{K}T_i^{k,l}(\pi_{i,k}-\pi'_{i,k})|\\
 &\leq \sum\limits_{i=1}^{L} \lambda_i \|\bm{\pi}_i - \bm{\pi}'_i\|_1\\
 &\leq \lambda_{max}\sum\limits_{i=1}^{L} \|\bm{\pi}_i - \bm{\pi}'_i\|_1\\
 &\leq \lambda_{max}\|\bm{\pi} - \bm{\pi}'\|_1,
\end{split}
\end{equation}where $\lambda_{max} = \max \{\lambda_1,\ldots,\lambda_L\}$.
As discussed above, $\lambda_{max}$ meets $0<\lambda_{max}<1$. Thus we can conclude that there exists a unique fixed point $\bm{\pi}^*$ and the iteration process ($\ref{iteration}$) can converge to $\bm{\pi}^*$.
\end{proof}
Moreover, the iteration process can stop if
\begin{equation}
\|\bm{\pi}^{(s)} - \bm{\pi}^{(s-1)}\| \leq e_{\pi},
\end{equation}where $e_{\pi}$ is a small positive value. Then, after setting the initial value of $\bm{\pi}^{(1)}$ and repeating the FI method, we can obtain the steady state distributions of all IoDs. Here, we would like to clarify that the FI algorithm is adopted to evaluate the system performance of the proposed scheduling policies, which is not required for the practical implementation of these policies at all.
\subsection{System Outage Probability and Average Throughput}
Based on the steady state analysis above, we now can evaluate the system outage probability and average throughput. In the proposed throughput-oriented policy, an outage occurs when all IoDs do not have enough energy in their batteries or the selected IoD transmits information but the received SINR at the HAP is less than the required threshold. Thus, the system outage probability can be expressed as
\begin{equation}
\label{FD_out}
P^{\mathrm{T}}_{out} = P^{\mathrm{T}}_0 + \sum\limits_{i=1}^{L}P^{\mathrm{T}}_{i,out}.
\end{equation}
In ($\ref{FD_out}$), $P^{\mathrm{T}}_0$ is the probability that all the IoDs have no enough energy in their batteries, which can be calculated as
\begin{equation}
\label{FD_p0}
P^{\mathrm{T}}_{0} = \prod\limits_{i=1}^{L}\pi_{i,0},
\end{equation}
and $P^{\mathrm{T}}_{i,out}$ is the probability of the event that the $i$-th IoD is selected to perform IT but an outage occurs. Based on the previous analysis, we can express this probability as a function of $\pi_{i,k}$ and $\upsilon^{\mathrm{T}}_{i,k}$, given by
\begin{equation}
\label{FD_iout}
\begin{split}
 &P^{\mathrm{T}}_{i,out} = \sum\limits_{k=1}^{K}\upsilon^{\mathrm{T}}_{i,k}\textrm{Pr}\bigg(\Big(P_i = \frac{\varepsilon_k}{T}\Big) \cap \Big(\log_2(1+\gamma_i)<R\Big)\bigg)\\
 &~~~~~~= \sum\limits_{k=1}^{K}\pi_{i,k}\upsilon^{\mathrm{T}}_{i,k}F_{H^U_i}\Big(\frac{K(2^{R}-1)(\alpha P_H+N_0)T}{kC}\Big).\\
 \end{split}
\end{equation}
Substituting ($\ref{FD_p0}$) and ($\ref{FD_iout}$) into ($\ref{FD_out}$), we have attained an analytical expression for the system outage probability of the considered FD WP-IoT system implementing the proposed throughput-oriented scheduling scheme.

We can then calculate the system average throughput $\Omega^{\mathrm{T}}$, given by
\begin{equation}\label{eq:rate_throughput}
\Omega^{\mathrm{T}} = R(1-P^{\mathrm{T}}_{out})T.
\end{equation}As $P_H$ increases, the battery steady state distribution of each IoD will present the trend that $\pi_{i,0}$ decreases and $\pi_{i,K}$ increases. For the term $\upsilon^\textrm{T}_{i,k}$, as we can observe from Eq. ($\ref{v_WRE_ik}$), the set $\mathcal{G}^{\mathrm{T}}_{p,k}$ only includes the battery states from 0 to $\frac{\varepsilon_k\bar{H}_i}{\bar{H}_p}$. When $P_H$ is small, the batteries of most of the IoDs are located at the low states and thus make $\upsilon^\textrm{T}_{i,k}$ large. When $P_H$ increases, the batteries of most of the IoDs are located at the high states and make $\upsilon^\textrm{T}_{i,k}$ smaller. As a result, as $P_H$ increases, $\upsilon^\textrm{T}_{i,k}$ will decrease.

\textit{Remark 1:} From the above analysis, we can see that the two terms in ($\ref{FD_out}$), $P^\textrm{T}_0$ and $\sum\limits_{i=1}^{L}P^{\mathrm{T}}_{i,out}$, reflect different trends as the HAP transmit power $P_H$ increases. More specifically, when $P_H$ is small, $P^\textrm{T}_0$ closes to 1 and $\sum\limits_{i=1}^{L}P^{\mathrm{T}}_{i,out}$ closes to 0. This is because in this case, each IoD cannot accumulate too much energy thus $P^\textrm{T}_0\rightarrow 1$ so that the system outage probability closes to 1. When $P_H$ increases, $P^\textrm{T}_0$ becomes smaller and the term $\sum\limits_{i=1}^{L}P^{\mathrm{T}}_{i,out}$ becomes larger. This is because more IoDs can accumulate enough energy to transmit data to the HAP and the data can be received by the HAP with high success probability due to the light effect of SI.  As a result, the system outage becomes smaller. If $P_H$ keeps increasing and becomes very large, $P^\textrm{T}_0$ will close to 0 because the battery of each IoD could always have enough energy for information transmission. Thus, within each time block, there always exists IoDs who can transmit data to the HAP. However, in this case, although HAP can receive data almost in each time block, the received SINR can be low due to the severe SI. This will lead to the fact that the term $\sum\limits_{i=1}^{L}P^{\mathrm{T}}_{i,out}$ closes to 1. In this case, the system outage probability will be high. Hence, the system outage probability will firstly decrease and then go up as $P_H$ increases and we can deduce that there should be an optimal value of $P_H$ to minimize the system outage probability and thus maximize the system average throughput.

\section{Design and Analysis of the Fairness-oriented Policy}
\subsection{Policy Design}
 Although the throughput-oriented policy proposed in the previous section can achieve good throughput performance, it can lead to the unfair issue among IoDs. Specifically, the IoDs with low average channel gains will almost have no chance to transmit data. To address this issue, the most widely used method in traditional communication systems is the normalized throughput based scheduling method [$\ref{PF1}$]. This method selects out the user with the largest ratio between instantaneous and average rate among all active users in the system. As such, this method reduces the probability that the scheduler always picks the user with the highest instantaneous rate, thus can effectively improve user fairness in a long term. In contrast to traditional communication systems, the IoD in the proposed FD WP-IoT system has no instantaneous CSI and is energy limited. The available information of each IoD is the current residual energy, the number of waiting time blocks from the latest data transmission and the statistical CSI. The current residual energy shows the exact accumulated energy during waiting time, which is defined as the time period from the latest IT operation to now. With the statistical CSI, the average accumulated energy during the waiting time can be calculated. Inspired by the normalized SNR-based method [$\ref{PF1}$], at the beginning of the $t$-th time block, we propose a fairness-oriented scheduling scheme that selects the $i$-th IoD if $i$ meets
\begin{equation}\label{PF_condition}
i = \arg \max\limits_{j\in\{1,2,\ldots,L\}}\Big\{\frac{r_j[t]}{W_j[t]\phi_j}\Big\}~\textrm{and}~r_i[t] \neq 0,
\end{equation}where $W_j[t]$ is the waiting time from the latest data transmission of the $j$-th IoD, and $\phi_j$ is the average harvested energy per block at $S_j$, which has the form
\begin{equation}\label{fai}
{\phi}_j = \eta P_H \bar{H}_jT.
\end{equation}Due to the limitation of the battery capacity, the average harvested energy during time period $W_i[t]$ is bounded, which is
\begin{equation}\label{W_harvested_energy}
W_i[t]{\phi}_i = \textrm{min} \Big\{W_i[t]{\phi}_i, C\Big\}.
\end{equation}
From ($\ref{PF_condition}$), we can see that the proposed fairness-oriented policy uses the normalized accumulated energy instead of its absolute value to schedule the IoDs.

 Based on the explanation above, we can also use the finite state MC to model the IoD's behaviors in fairness-oriented case. However, compared ($\ref{PF_condition}$) with ($\ref{condition}$), in the fairness-oriented case, besides the residual energy, the number of waiting blocks should also be treated as the system state. In this sense, the one state MC model presented in Sec. III.B is no longer applicable to analyze the performance of the proposed fairness-oriented method. A new framework is thus needed for constructing the state transition matrix, which is elaborated in detail in the subsequent subsection.
\subsection{MC Formulation and Steady State Calculation}
  We let $\frac{k C}{Km\phi_i}$ denote the discrete state that $S_i$ locates at energy level $k$ ($0\leq k \leq K$) while it has waited $m$ time blocks ($m \geq 1$) since its last scheduled transmission. To describe the state transition, we use $T^{(k,m),(l,u)}_i$ to denote the probability of the transition from state $(k,m)$ (initial energy level $k$ and waiting time blocks $m$) to state $(l,u)$ (end energy level $l$ and waiting time block $u$). Theoretically, the number of waiting blocks $m$ can be as large as infinity. This case will make the IoDs' state transition matrix intractable. To cope with this issue, we impose a reasonable and practical assumption: each IoD holds a maximum waiting time $M$. If $W_i[t] < M$, $S_i$ will be scheduled to transmit data only when ($\ref{PF_condition}$) holds. On the other hand, if $W_i[t] = M$ and $r_i[t]>0$, $S_i$ will be scheduled to transmit data immediately without considering the states of other IoDs. However, if there exists any other IoDs which have waited $M$ time blocks as well, the scheduler will select the IoD which holds the maximum weighted residual energy. Once $S_i$ is scheduled to transmit data at the $t$-th time block, at the beginning of the next time block, its waiting time will be initialized to 1. For the case $W_i[t] = M$ and $r_i[t] = 0$, $S_i$ will hold its waiting time and continue to harvest energy. Mathematically, at the beginning of the $t$-th time block, the $i$-th IoD will be selected in the fairness-oriented policy if $i$ meets
  \begin{small}
\begin{equation}
\label{FD_PF_condition2}
i=
\begin{cases}
\arg \max\limits_{j\in\{1,2,\ldots,L\}}\Big\{\frac{r_j[t]}{W_j[t]\phi_j}\Big\} ,~\textrm{if}~\chi = \emptyset~\textrm{and}~r_i[t]\neq 0\\
\arg\max\limits_{j\in\chi}\Big\{r_j[t]\bar{H}_j\Big\} ,~\textrm{if}~\chi \neq \emptyset~\textrm{and}~r_i[t]\neq 0\\
\end{cases},
\end{equation}\end{small}where $\chi$ denotes the set including all the IoDs of which waiting time blocks equal to the threshold $M$.
We thus re-write ($\ref{FD_states}$) as
\begin{equation}
\label{FD_PF_states}
\zeta_i[t]=
\begin{cases}
\zeta_T,~\textrm{if}~(\ref{FD_PF_condition2})~\textrm{holds}\\
\zeta_E,~\textrm{otherwise}\\
\end{cases}.
\end{equation}The evolution of $r_i[t]$ from $t$ to $t+1$ can be expressed as
\begin{equation}
\label{E_PF_changed}
r_i[t+1]=
\begin{cases}
0,~\textrm{if}~\zeta_i[t]=\zeta_T\\
\textrm{min}\Big\{r_i[t]+E_i, C\Big\}, \textrm{if}~\zeta_i[t]=\zeta_E\\
\end{cases}
,
\end{equation}and that of $W_i[t]$ from $t$ to $t+1$ is expressed as
\begin{equation}
\label{T_PF_changed}
W_i[t+1]=
\begin{cases}
1,~\textrm{if}~\zeta_i[t]=\zeta_T\\
\textrm{min}\Big\{W_i[t]+1, M\Big\}, \textrm{if}~\zeta_i[t]=\zeta_E\\
\end{cases}
.
\end{equation}
Similar to the throughput-oriented scheme, we can also summarize the state transition probabilities of the MC for each IoD into \textit{eight} cases depending on different initial energy level $k$. On the other hand, as ($\ref{FD_PF_condition2}$) shows, in the fairness-oriented policy the priority of data transmissions depends on not only the initial energy level, but also the number of waiting time blocks. Thus, for different initial waiting blocks $m$ and end waiting blocks $u$, we can further divide the state transition probability into $four$ cases: 1) The waiting time block increases by one; 2) The waiting time block returns to $1$ when $m < M$; 3) The waiting time block remains unchanged at $M$; 4) The waiting time block returns to $1$ when $m = M$. We thus have in total 32 system states by joint considering the energy status and waiting time status ($k,m$). Mathematically, the transition probability of the $i$-th IoD from state ($k,m$) to state ($l,u$) can be expressed as:

\textit{1) $u=m+1$ when $m < M$:} It describes the case that $S_i$ is not chosen to transmit data while $W_i[t] < M$. Mathematically, the transition probability of this general can be expressed as
\begin{small}
\begin{equation}\label{PF_T_case_1}
T^{(k,m),(l,m+1)}_i=
\begin{cases}
\textrm{Pr}(\tilde{E_i} \leq \frac{C}{K}),~\textrm{if}~k = l=0\\
\textrm{Pr}\Big(\frac{(l-k)C}{K}\leq \tilde{E_i} \leq \frac{(l-k+1)C}{K}\Big),\\
~~~~~~~~~~~~~~~~~~~~~\textrm{if}~k < l < K,~k=0\\
   \textrm{Pr}\Big(\tilde{E_i} \geq \frac{(K-k)C}{K}\Big),~\textrm{if}~k=0,~l = K\\
   0,~\textrm{if}~k > l,~l>0\\
   0,~\textrm{if}~k>0,~l=0\\
   (1-\upsilon^{\mathrm{F}}_{i,(k,m)})\textrm{Pr}(\tilde{E_i} \leq \frac{C}{K}),\\
   ~~~~~~~~~~~~~~~~~~~~\textrm{if}~k >0,~k = l \neq K\\
   (1-\upsilon^{\mathrm{F}}_{i,(k,m)})\textrm{Pr}\Big(\frac{(l-k)C}{K}\leq \tilde{E_i} \leq \frac{(l-k+1)C}{K}\Big),\\
   ~~~~~~~~~~~~~~~~~~~\textrm{if}~0<k < l \neq K\\
   (1-\upsilon^{\mathrm{F}}_{i,(k,m)})\textrm{Pr}(\tilde{E_i} \geq \frac{(K-k)C}{K}),\\
   ~~~~~~~~~~~~~~~~~~~~\textrm{if}~k>0,~l = K
\end{cases}
,
\end{equation}
\end{small}where $\upsilon^{\mathrm{F}}_{i,(k,m)}$ represents the probability that the $i$-th IoD is scheduled to operate in the IT mode at battery level $k$ and waiting time blocks $m$.

\textit{2) $u=1$ when $m < M$:} It describes the case that $S_i$ is scheduled to transmit data while its waiting period is below the threshold. In this case, the end state has only one possibility, which can be expressed as
\begin{equation}\label{PF_T_case_2}
T^{(k,m),(l,1)}_i=
\begin{cases}
   \upsilon^{\mathrm{F}}_{i,(k,m)},~\textrm{if}~k>0,~l=0\\
   0,~\textrm{otherwise}\\
\end{cases}
.
\end{equation}

\textit{3) $u=m$ when $m = M$:} It describes the case that $S_i$ is not scheduled to transmit data when its waiting block number is $M$. This case happens when $S_i$ holds no available energy, or there exists other IoD which has waited $M$ time blocks. Hence, the transition probability of this case can be expressed as
\begin{small}
\begin{equation}\label{PF_T_case_3}
T^{(k,M),(l,M)}_i=
\begin{cases}
\textrm{Pr}(\tilde{E_i} \leq \frac{C}{K}),~\textrm{if}~k = l=0\\
\textrm{Pr}\Big(\frac{(l-k)C}{K}\leq \tilde{E_i} \leq \frac{(l-k+1)C}{K}\Big),\\
~~~~~~~~~~~~~~~~~~~~~\textrm{if}~k < l < K,~k=0\\
   \textrm{Pr}\Big(\tilde{E_i} \geq \frac{(K-k)C}{K}\Big),~\textrm{if}~k=0,~l = K\\
   0,~\textrm{if}~k > l,~l>0\\
   0,~\textrm{if}~k>0,~l=0\\
   (1-\upsilon^{\mathrm{F}}_{i,(k,M)})\textrm{Pr}(\tilde{E_i} \leq \frac{C}{K}),\\
   ~~~~~~~~~~~~~~~~~~~~~\textrm{if}~k >0,~k = l \neq K\\
   (1-\upsilon^{\mathrm{F}}_{i,(k,M)})\textrm{Pr}\Big(\frac{(l-k)C}{K}\leq \tilde{E_i} \leq \frac{(l-k+1)C}{K}\Big),\\
   ~~~~~~~~~~~~~~~~~~~~~\textrm{if}~0<k < l \neq K\\
   (1-\upsilon^{\mathrm{F}}_{i,(k,M)})\textrm{Pr}(\tilde{E_i} \geq \frac{(K-k)C}{K}),\\
   ~~~~~~~~~~~~~~~~~~~~~~\textrm{if}~k>0,~l = K
\end{cases}
.
\end{equation}
\end{small}

\textit{4) $u=1$ when $m = M$:} It describes the case that $S_i$ is scheduled to transmit data when its number of waiting blocks is $M$. The transition probability of this case can be expressed as
\begin{equation}\label{PF_T_case_4}
T^{(k,M),(l,1)}_i=
\begin{cases}
   \upsilon^{\mathrm{F}}_{i,(k,M)},~\textrm{if}~k>0,~l=0\\
   0,~\textrm{otherwise}\\
\end{cases}
.
\end{equation}The transition probability of other cases is equal to $0$.

Before calculating $\upsilon^{\mathrm{F}}_{i,(k,m)}$, we use a vector $\bm{\pi}_i$
to express the steady state distribution in different residual energy and the number of waited blocks of $S_i$, where the expression of $\bm{\pi}_i$ is given in ($\ref{pi_i}$) on top of the next page. Before calculating these steady states, we first derive the relationship between $\upsilon^{\mathrm{F}}_{i,(k,m)}$ and the stationary state of all IoDs' battery.
\begin{figure*}[ht]
\vspace*{3pt}
\normalsize
\setcounter{MYtempeqncnt}{\value{equation}}
\begin{equation}\label{pi_i}
\bm{\pi}_i =[{\pi}_{i,(0,1)},\cdots,{\pi}_{i,(0,M)},\cdots,{\pi}_{i,(k,1)},\cdots,{\pi}_{i,(k,M)},\cdots,{\pi}_{i,(K,1)},\cdots,{\pi}_{i,(K,M)}]^T
\end{equation}
\hrulefill
\vspace*{3pt}
\end{figure*}

Recall the scheduling principle mentioned above, when $S_i$'s waiting time blocks $m$ is below the threshold $M$, the event that $S_i$ operates in the IT mode at the battery level $k$ happens only when its normalized accumulated energy is the maximum. In this case, $\upsilon^{\mathrm{F}}_{i,(k,m)}$ can be expressed as
\begin{equation}
\label{v_PF_k/m_m<M}
\upsilon^{\textrm{F}}_{i,(k,m)} =\prod\limits_{p\in\varphi\setminus\{i\}}\Big(\sum\limits_{(q,u)\in\mathcal{G}^{\textrm{F}}_{p,(k,m)}}\pi_{p,(q,u)}\Big), ~\textrm{if}~k \geq 0,~m < M,
\end{equation}where $\mathcal{G}^{\textrm{F}}_{p,(k,m)}$ is the set including all the states of the $p$-th IoD that meet $\frac{q}{u}<\frac{k\phi_p}{m\phi_i},~\forall~(q,u) \in \mathcal{G}^{\textrm{F}}_{p,(k,m)}$. When $m = M$, $S_i$ will always be scheduled to transmit data, unless there exists other IoD who has waited $M$ time blocks as well and has a larger weighted residual energy. In this case, $\upsilon^{\textrm{F}}_{i,(k,m)}$ can be expressed as
\begin{equation}
\label{v_PF_k/m_m=M}
\begin{split}
\upsilon^{\textrm{F}}_{i,(k,m)} &=\prod\limits_{p\in\varphi\setminus\{i\}}\Big(\sum\limits_{u\in[1,M-1]}\sum\limits_{q\in[0,K]}\pi_{p,(q,u)} \\
&+ \sum\limits_{l\in\mathcal{G}^\textrm{F}_{p,k}}\pi_{p,(l,M)}\Big),\\
&~\textrm{if}~k \geq 0,~m = M,\\
\end{split}
\end{equation}where $\mathcal{G}^\textrm{F}_{p,k}$ is the set including all the battery levels of $p$-th IoD that smaller than $\frac{\varepsilon_k \bar{H}_i}{\bar{H}_p}$ in fairness-oriented scheme. Similar to the previous section, the steady state for each IoD's battery and discrete waiting time blocks can also be expressed as a equation set like ($\ref{PF1}$). Thus, the IoDs' steady state for the fairness-oriented scheme can be calculated through the FI method like ($\ref{iteration}$).
%
%
%
\subsection{Performance Evaluation}
Similar with the throughput-oriented scheme, the outage probability and system average throughput in the fairness-oriented scheduling scheme can be expressed as
\begin{equation}\label{eq:fair_outage}
P^{\textrm{F}}_{out} = P^{\textrm{F}}_0 + \sum\limits_{i=1}^LP^{\textrm{F}}_{i,out},
\end{equation}
\begin{equation}\label{eq:fair_throughput}
\Omega^{\mathrm{F}} = R(1-P^{\mathrm{F}}_{out})T,
\end{equation}where
\begin{equation*}
P^{\textrm{F}}_0 = \prod\limits_{i=1}^L \Big(\sum\limits_{m=1}^M\pi_{i,(0,m)}\Big)
\end{equation*}denotes the probability that all the IoDs have no available stored energy. $P^{\textrm{F}}_{i,out}$ denotes the probability that $S_i$ is scheduled to transmit data but outage happens. Different from $P^{\textrm{T}}_{i,out}$ defined in ($\ref{FD_iout}$), $P^{\textrm{F}}_{i,out}$ needs to consider not only the energy level but also the waiting time blocks, which has the form
\begin{small}
\begin{equation}\label{p_i_out_fairness}
\begin{split}
 P^{\mathrm{F}}_{i,out} &= \sum\limits_{k=1}^{K}\sum\limits_{m=1}^{M}\upsilon^{\mathrm{F}}_{i,(k,m)}\\
 &\times\textrm{Pr}\bigg(\Big(P_i = \frac{\varepsilon_k}{T}\Big)\cap \Big(W_i[t] = m\Big) \cap \Big(\log_2(1+\gamma_i)<R\Big)\bigg)\\
 &= \sum\limits_{k=1}^{K}\sum\limits_{m=1}^{M}\pi_{i,(k,m)}\upsilon^{\mathrm{F}}_{i,(k,m)}F_{H^U_i}\Big(\frac{K(2^{R}-1)(\alpha P_H+N_0)T}{kC}\Big).\\
 \end{split}
\end{equation}
\end{small}
To calculate the time difference (measured by rounds of transmission blocks) between two adjacent data transmissions, we consider a sufficient long period. Denote by $N$ the total number of transmission blocks that the system undergoes with $N \rightarrow \infty$, For a certain IoD $S_i$, the total number of transmission blocks $N$ can be divided into two groups: $N^T_i$-number of information transmission blocks and $N^E_i$-number of energy harvesting blocks, where $N=N^T_i+N^E_i$. Recall that the probability that the $i$-th IoD stays at the battery level $k$ waiting time block $m$ is $\pi_{i,(k,m)}$ and the probability that the $i$-th IoD is scheduled to transmit information at battery level $k$ and waiting time block $m$ in fairness-oriented scheme is $\upsilon^\textrm{F}_{i,(k,m)}$. We thus can calculate the probability that the $i$-th IoD is scheduled to operate in the IT mode under the fairness-oriented scheme, given by
\begin{equation}
\upsilon^\textrm{F}_i = \sum\limits_{k=0}^{K}\sum\limits_{m=1}^{M}\pi_{i,(k,m)}\upsilon^\textrm{F}_{i,(k,m)}.
\end{equation}Similarly, the selection probability of $S_i$ under throughput-oriented scheme is given by
\begin{equation}
\upsilon^\textrm{T}_i = \sum\limits_{k=0}^{K}\pi_{i,k}\upsilon^\textrm{T}_{i,k}.
\end{equation}
We then have $N^T_i = \upsilon^\textrm{F}_i N$ and $N^E_i = (1-\upsilon^\textrm{F}_i)N$ under fairness-oriented scheme ($N^T_i = \upsilon^\textrm{T}_i N$ and $N^E_i = (1-\upsilon^\textrm{T}F_i)N$ under throughput-oriented scheme). Therefore, the average rounds required to charge battery between two successive information transmission for the $i$-th IoD in different scheduling schemes can be expressed as\begin{equation}
\overline{T}_i = \frac{N^E_i}{N^T_i} = \frac{(1-\upsilon_i)N}{\upsilon_i N} = \frac{1-\upsilon_i}{\upsilon_i}, \upsilon_i\in [\upsilon^\textrm{T}_i, \upsilon^\textrm{F}_i].
\end{equation}In the next section, we will validate the analysis above through the simulation results.

\section{Numerical Results and Discussion}
In this section, we validate the analytical expressions derived in the previous sections by simulations. In order to capture the effect of path loss on the system performance, we set $\bar{H}_i=\frac{1}{1+d^\epsilon_i}$, where $d_i$ is the distance between $S_i$ and HAP, and $\epsilon \in [2,5]$ is the path loss exponent. In all simulations, we set the noise power $N_0=-60$dBm, the energy conversion efficiency $\eta=0.5$, the battery capacity $C = 5 \times 10^{-4}$ Joule, Rician factor $\Psi = 6$ which is typical for a line-of-sight indoor environment [$\ref{Rician_index}$], the path loss exponent $\epsilon = 3$, loop interference channel power gain $\alpha=10^{-7}$ and the maximum battery level $K=200$. The total number of IoDs $L$ is 5. For the fairness-oriented scheme, the maximum number of waiting time blocks $M$ is set to $50$.

To show the effectiveness of the proposed scheduling policies, a comparison of the proposed scheduling schemes with two other policies is provided. The first benchmark scheme is the round robin (RR) scheme. In the RR scheme, each IoD is scheduled in a fixed order without the consideration of either the channel quality or the residual energy. Another scheme is a random selection (RS) scheme. In the RS scheme, each IoD within the system has the same probability to access the channel, and at each transmission block, the scheduler selects an IoD randomly.

\begin{figure}[htbp]
\includegraphics[height=2.5in]{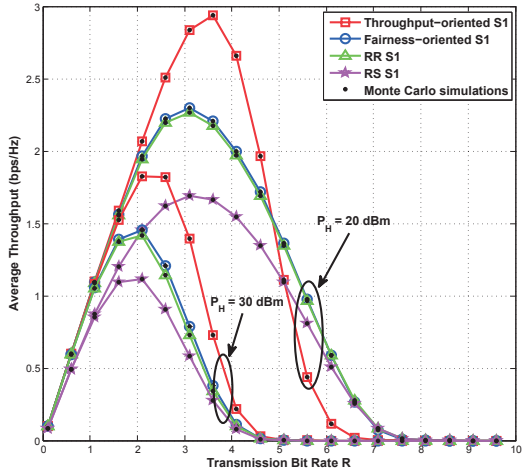}
\centering
\caption{\label{Fig:thput_1}System average throughput versus transmission bit rate $R$ with system setup $S1$: $d_1 = 8$m, $d_2 = 9$m, $d_3 = 10$m, $d_4 = 11$m and $d_5 = 12$m.}
\end{figure}
\begin{figure}[htbp]
\includegraphics[height=2.5in]{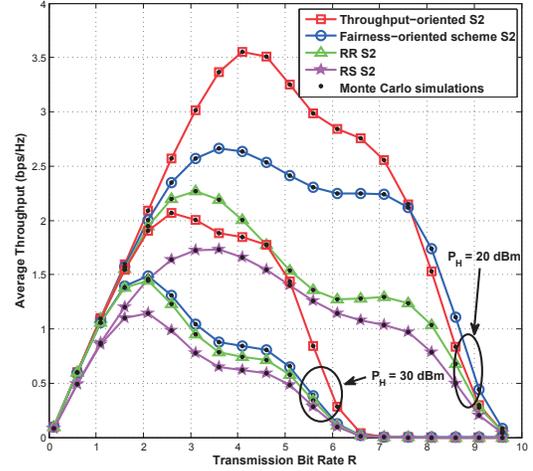}
\centering
\caption{\label{Fig:thput_2}System average throughput versus transmission bit rate $R$ with system setup $S2$: $d_1 = 5$m, $d_2 = 9$m, $d_3 = 10$m, $d_4 = 11$m and $d_5 = 12$m.}
\end{figure}

Figs. $\ref{Fig:thput_1}$ and $\ref{Fig:thput_2}$ compare the system average throughput of  different scheduling schemes mentioned above for different rate requirement $R$ and different system setups. We can see that the analytical expressions derived in ($\ref{eq:rate_throughput}$) and ($\ref{eq:fair_throughput}$) agree well with the corresponding Monte Carlo simulations, which validates our theoretical analysis. It can also be observed that for a given system setup, there exists a maximum system average throughput $\Omega_{max}$. On the other hand, when $R$ is high enough, the throughput-oriented scheme will present the worst throughput performance among all the schemes. This is because for the throughput-oriented scheme, due to the adopt of the weighted residual energy as selection criterion, the IoDs with good channel quality get most of the access opportunity and contribute the system throughput mostly. However, these frequently chosen IoDs could have limited time to accumulate energy such that the maximum rate they can afford is low. Thus, once $R$ is beyond the maximum capacity that these IoDs can support, the system outage probability will increase dramatically, which leads to a low system throughput. Different from the throughput-oriented scheme, the IoDs of the other schemes share the access opportunity more fairly, which means the IoDs with good channel quality have longer time to accumulate energy and thus can afford higher rate requirement. As a consequence, the throughput-oriented presents the worst throughput performance in large $R$ value. In Fig. $\ref{Fig:thput_1}$, the throughput improvement of our proposed fairness-oriented throughput scheme is not significant compared with that of the RR scheme. However, for an alternative system setup with more diverse channel variances, as shown in Fig. $\ref{Fig:thput_2}$, the performance gap between the proposed fairness-oriented scheme and the RR scheme is noteworthy. The reason is as follow. Firstly, let us recall the definition of the normalized residual energy $\frac{r_i[t]}{W_i[t]\phi_i}$. For each IoD, it is straightforward that the mean value of the normalized residual energy should be 1. In the system whose channel variances are diverse, the increment of the normalized residual energy per block of different IoDs could be very diverse. This may further lead to the fact that the normalized residual energy of different IoDs are very diverse. In this case, some IoDs will hold the normalized residual energy far from their mean value 1 more easily compared with the system whose channel variances is not so diverse. Secondly, recall the definition of the fairness-oriented policy, if the normalized-residual energy of a certain IoD is much higher than its average value, it means this IoD has harvested much more energy than its  average amount of harvested energy and this IoD will be chosen to transmit data with high probability. In contrast with the fairness-oriented policy, the RR policy only schedules each IoD in the fixed order and does not take advantage of this characteristic. This may lead a fact that the IoD scheduled to transmit data just holds few normalized residual energy. For this reason, the performance gap between the fairness-oriented policy and the RR policy is noteworthy for an alternative system setup with more diverse channel variances. As such, Figs. $\ref{Fig:thput_1}$ and $\ref{Fig:thput_2}$ show different performance gaps between the fairness-oriented scheme and the RR scheme.

\begin{figure}[htbp]
\includegraphics[height=2.6in]{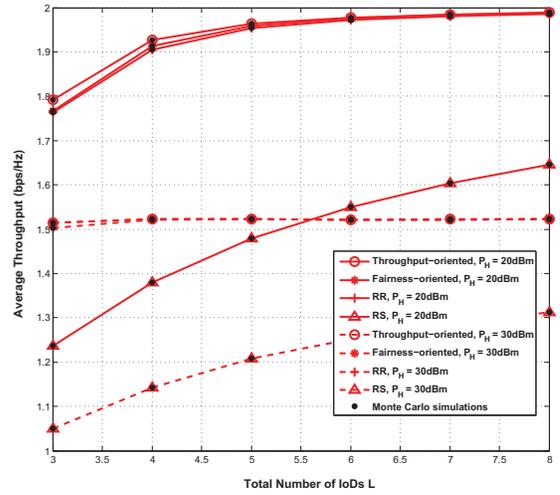}
\centering
\caption{\label{Fig:thput_iid}System average throughput versus number of IoDs with $R$ = 2. With different total number of IoDs $L$, $d_i$ = 10m, $\forall i$, $i\in\{1,2,\ldots,L\}$.}
\end{figure}

\begin{figure}[htbp]
\includegraphics[height=2.6in]{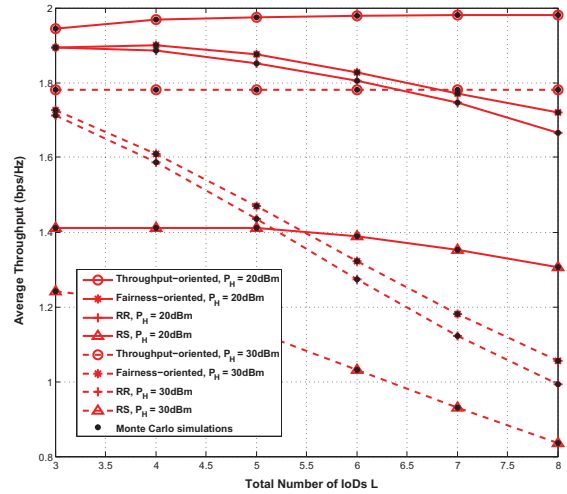}
\centering
\caption{\label{Fig:thput_diff}System average throughput versus number of IoDs with $R$ = 2. The system setup is: with different total number of IoDs $L$, $d_i$ = $(7+i)$m, $i\in\{1,2,\ldots,L\}$.}
\end{figure}

To investigate the effect of the IoD number on system performance, we depict the system average throughput versus the total number of IoDs in different system setups through Figs. $\ref{Fig:thput_iid}$ and $\ref{Fig:thput_diff}$. In Fig. $\ref{Fig:thput_iid}$, the distance between each IoD and HAP is set to 10m. In Fig. $\ref{Fig:thput_diff}$, with different total number of IoDs L, the distance between $i$-th IoD and HAP is set: $d_i = (7+i)$m. From Figs. $\ref{Fig:thput_iid}$ and $\ref{Fig:thput_diff}$, we can find as the distance between each IoD and HAP becomes more diverse, the performance gap of the system average throughput between fairness-oriented policy and RR is becoming more distinct. This validates our guess proposed above.

\begin{figure}[htbp]
\includegraphics[height=2.5in]{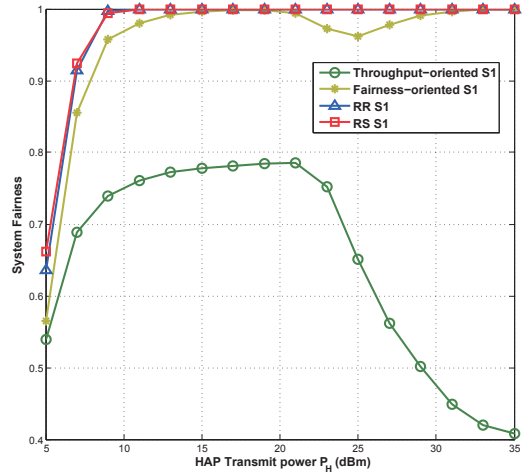}
\centering
\caption{\label{Fig:fairness}System fairness versus HAP transmit power with system setup $S1$: $d_1 = 8$m, $d_2 = 9$m, $d_3 = 10$m, $d_4 = 11$m and $d_5 = 12$m.}
\end{figure}

\begin{figure}[htbp]
\includegraphics[height=2.5in]{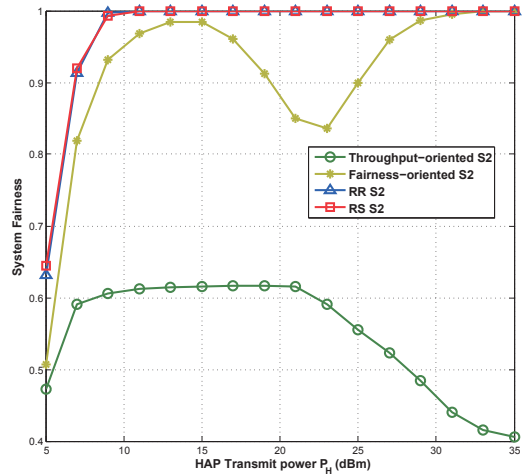}
\centering
\caption{\label{Fig:fairness2}System fairness versus HAP transmit power with system setup $S2$: $d_1 = 5$m, $d_2 = 9$m, $d_3 = 10$m, $d_4 = 11$m and $d_5 = 12$m.}
\end{figure}

Figs. $\ref{Fig:fairness}$ and $\ref{Fig:fairness2}$ compare the system average fairness of the proposed scheduling policies and their benchmarks. To quantify how fair the system resource is allocated among all IoDs, in this paper we adopt the notion developed in [$\ref{fairness_define}$] to evaluate the system fairness performance. As [$\ref{fairness_define}$] explained, the average fairness of a system can be defined as
\begin{equation}\label{eq:def_fair}
\bar{f} = -\sum\limits_{i=1}^L\rho_i \frac{\log(\rho_i)}{\log(L)},
\end{equation}where $\rho_i$ is the access probability of $S_i$. The access probability of the fairness-oriented scheme $\rho^\textrm{F}_i$ can be calculated as
\begin{equation}\label{eq:PF_access_probability}
\rho^\textrm{F}_i = \sum\limits_{k=1}^K \sum\limits_{m=1}^M\pi_{i,(k,m)} \upsilon^\textrm{F}_{i,(k,m)}.
\end{equation}The access probability of the throughput-oriented scheme $\rho^\textrm{T}_i$ can be expressed as
\begin{equation}\label{eq:T_access_probability}
\rho^\textrm{T}_i = \sum\limits_{k=1}^K \pi_{i,k} \upsilon^\textrm{T}_{i,k}.
\end{equation}
As shown in ($\ref{eq:def_fair}$), the system will achieve better fairness if the IoDs share the resource more fairly.

 We can see from Figs. $\ref{Fig:fairness}$ and $\ref{Fig:fairness2}$ that the fairness-oriented scheme can achieve almost the same fairness performance as the RR and RS, while the throughput-oriented scheme achieves poor fairness performance. As $P_H$ increases, the fairness of the throughput-oriented scheme firstly increases and then decreases. This is because When $P_H$ is small, all schemes demonstrate bad fairness since the far IoDs cannot harvest any energy so that they almost have no opportunity to transmit data. When $P_H$ is extremely large, at the beginning of each time block,  the battery of every IoD is always full. Meanwhile, under the situation that every IoD holds the same residual energy, due to the weighting effect of the average channel power gain, the IoDs nearer to the HAP will possess higher weighted residual energy and thus dominate the access process. Thus, the system fairness reduces. When it comes to the fairness-oriented scheme, as the HAP transmit power increases, the fairness has a local minimum value. To explain this phenomenon, let us review the node selection criterion of the fairness-oriented scheme given in ($\ref{PF_condition}$). From ($\ref{PF_condition}$), we can see that the system will schedule the $i^*$-th IoD to transmit data when it has the largest normalized harvested energy $
\frac{r_i[t]}{W_i[t]\phi_i}$ among all the IoDs. Moreover, from ($\ref{W_harvested_energy}$), as $P_H$ increases, the denominator of the normalized harvested energy cannot keep increasing towards infinitely and will be truncated at $C$. Otherwise, as ($\ref{fai}$) shows, the IoD closest to the HAP, whose average channel quality is the highest, will meet the condition above firstly. In this sense, as $P_H$ becomes larger, the increasing speed of the normalized harvested energy of this closest IoD will be faster than the other IoDs such that it will obtain more access opportunities and the system fairness will decline. On the other hand, as $P_H$ keeps increasing, the system fairness will climb up again. This is because when $P_H$ is large enough, not only the denominator, the numerator of the normalized harvested energy will be bounded by the battery capacity $C$ as well, which makes the normalized harvested energy of all the IoDs close to 1. In this case, each IoD will have the same access probability and the system will thus demonstrate better fairness performance. In addition, if the difference of the average channel quality among different IoDs is enlarged, the phenomenon mentioned above will be more significant, as depicted in Fig. $\ref{Fig:fairness2}$. To find some ways to improve the local minimum fairness in the fairness-oriented scheme, we need to find the reason leading this case. The reason is that the average harvested energy is bounded by the battery capacity. Based on these observations, to improve the local minimum fairness, when any IoD detects that its average harvested energy is bounded, this IoD can introduce an increment to the average harvested energy to limit the growth rate of the normalized residual energy. In this way, the increasing speed of the normalized harvested energy of this IoD will be slowed down. This method can avoid the problem of unbalanced increasing speed of the denominator and numerator in $\frac{r_i[t]}{W_i[t]\phi_i}$ as $P_H$ increases. However, this increment should be carefully designed so that it will not over run.

\section{Conclusions}
In this paper, we proposed and evaluated two multi-user scheduling schemes for a full-duplex wireless-powered Internet-of-Things (WP-IoT) system. One is throughput-oriented, the other one is fairness-oriented. The charging/discharging processes of the battery of each wireless-powered node was modeled as a finite-state Markov chain. The analytical expression of the system outage probability and average throughput was derived for the proposed system over Rician fading channels. Simulation results were provided to validate our theoretical analysis. Comparisons of the proposed schemes with two other policies, round robin (RR) and random selection (RS) were also provided. Numerical results showed that the proposed throughput-oriented scheme presents the highest system average throughput. The proposed fairness-oriented scheme can achieve substantial throughput improvement over the RR and RS schemes while maintains good system fairness, especially when the average channel quality disparity among IoDs is larger. In our future work, we will consider the imperfect channel estimation and evaluate the system performance in finite blocklength regime.


\begin{thebibliography}{99}
\bibitem{1}\label{IOT_EH1}M. R. Palattella et al., "Internet of Things in the 5G Era: Enablers, Architecture, and Business Models," in \textit{IEEE Journal on Selected Areas in Communications}, vol. 34, no. 3, pp. 510-527, March 2016.
\bibitem{2}\label{IOT_EH2}P. Kamalinejad, C. Mahapatra, Z. Sheng, S. Mirabbasi, V. C. M. Leung and Y. L. Guan, "Wireless energy harvesting for the Internet of Things," in \textit{IEEE Communications Magazine}, vol. 53, no. 6, pp. 102-108, June 2015.
\bibitem{3}\label{IOT_EH3}D. Jayakody, J. Thompson, S. Chatzinotas, and S. Durrani, "Wireless Information and Power Transfer: A New Green Communications Paradigm", in \textit{Springer}, New York, USA, April, 2018.
\bibitem{4}\label{IOT_EH4}D. Jayakody, B. Chen, V. Sharma and K. Srinivasan, "Opportunistic Wireless Power Transfer Scheme for Multiple Access Relay Networks", in \textit{IEEE Access}, Aug, 2017.
\bibitem{small_cell}\label{Small_cell}J. Jang et al., "Smart Small Cell with Hybrid Beamforming for 5G: Theoretical Feasibility and Prototype Results," in \textit{IEEE Wireless Communications}, vol. 23, no. 6, pp. 124-131, Dec. 2016.
\bibitem{massive_MIMO}\label{massive_MIMO}E. G. Larsson, O. Edfors, F. Tufvesson and T. L. Marzetta, "Massive MIMO for next generation wireless systems," in \textit{IEEE Communications Magazine}, vol. 52, no. 2, pp. 186-195, Feb. 2014.
\bibitem{mmWave}\label{mmWave}P. Wang, Y. Li, L. Song and B. Vucetic, "Multi-gigabit millimeter wave wireless communications for 5G: from fixed access to cellular networks," in \textit{IEEE Communications Magazine}, vol. 53, no. 1, pp. 168-178, Jan. 2015.
\bibitem{low_power}\label{low_power}A. M. Zungeru, L. M. Ang, S. Prabaharan, and K. P. Seng, "Radio frequency energy harvesting and management for wireless sensor networks," \textit{Green Mobile Devices Netw.: Energy Opt. Scav. Tech.}, CRC Press, pp. 341-368, 2012
\bibitem{Jr2}\label{Introduction_WPCN}S. Bi, Y. Zeng and R. Zhang, "Wireless powered communication networks: an overview," in \textit{IEEE Wireless Communications}, vol. 23, no. 2, pp. 10-18, April 2016.
\bibitem{Jr3}\label{WPCN-Rui}H. Ju and R. Zhang, "Throughput Maximization in Wireless Powered Communication Networks," in \textit{IEEE Transactions on Wireless Communications}, vol. 13, no. 1, pp. 418-428, Jan. 2014.
\bibitem{Jr4}\label{WPCN-Rui2}L. Liu, R. Zhang and K. C. Chua, "Multi-Antenna Wireless Powered Communication With Energy Beamforming," in \textit{IEEE Transactions on Communications}, vol. 62, no. 12, pp. 4349-4361, Dec. 2014.
\bibitem{Jr5}\label{WPCN-PB}Y. Ma, H. Chen, Z. Lin, Y. Li and B. Vucetic, "Distributed and Optimal Resource Allocation for Power Beacon-Assisted Wireless-Powered Communications," in \textit{IEEE Transactions on Communications}, vol. 63, no. 10, pp. 3569-3583, Oct. 2015.
\bibitem{Jr6}\label{WPCN-HTC}H. Chen, Y. Li, J. L. Rebelatto, B. F. Uch?a-Filho and B. Vucetic, "Harvest-Then-Cooperate: Wireless-Powered Cooperative Communications," in \textit{IEEE Transactions on Signal Processing}, vol. 63, no. 7, pp. 1700-1711, April1, 2015.
\bibitem{URLLC1}\label{URLLC1}O. L. A. L\'{o}pez, H. Alves, R. D. Souza and E. M. G. Fern\'{a}ndez, "Ultrareliable Short-Packet Communications With Wireless Energy Transfer," in \textit{IEEE Signal Processing Letters}, vol. 24, no. 4, pp. 387-391, April 2017.
\bibitem{URLLC2}\label{URLLC2}O. L. A. L\'{o}pez, E. M. G. Fern\'{a}ndez, R. D. Souza and H. Alves, "Wireless Powered Communications with Finite Battery and Finite Blocklength," in \textit{IEEE Transactions on Communications}, vol. PP, no. 99, pp. 1-1, Dec. 2017.
\bibitem{URLLC3}\label{URLLC3}O. L. A. L\'{o}pez, E. M. G. Fern\'{a}ndez, R. D. Souza and H. Alves, "Ultra-Reliable Cooperative Short-Packet Communications With Wireless Energy Transfer," in \textit{IEEE Sensors Journal}, vol. 18, no. 5, pp. 2161-2177, March, 2018.
\bibitem{FD1}\label{FD_introduction}Y. Sun, D. W. K. Ng, Z. Ding and R. Schober, "Optimal Joint Power and Subcarrier Allocation for Full-Duplex Multicarrier Non-Orthogonal Multiple Access Systems," in \textit{IEEE Transactions on Communications}, vol.PP, no.99, pp.1-1, Jan. 2017.
\bibitem{33}\label{FD_WPCN_BF}M. Mohammadi, B. K. Chalise, H. A. Suraweera, C. Zhong, G. Zheng and I. Krikidis, "Throughput analysis and optimization of wireless-powered multiple antenna full-duplex relay systems," in \textit{IEEE Transactions on Communications}, vol. 64, pp. 1769-1785, Apr. 2016.
\bibitem{Jr7}\label{FD-WPCN-Rui}H. Ju and R. Zhang, "Optimal Resource Allocation in Full-Duplex Wireless-Powered Communication Network," in \textit{IEEE Transactions on Communications}, vol. 62, no. 10, pp. 3528-3540, Oct. 2014.
\bibitem{Jr8}\label{FD-WPCN-Sun}X. Kang, C. K. Ho and S. Sun, "Full-Duplex Wireless-Powered Communication Network With Energy Causality," in \textit{IEEE Transactions on Wireless Communications}, vol. 14, no. 10, pp. 5539-5551, Oct. 2015.
\bibitem{Jr9}\label{FD-WPCN-Sun2}T. P. Do and Y. H. Kim, "Resource Allocation for a Full-Duplex Wireless-Powered Communication Network With Imperfect Self-Interference Cancelation," in \textit{IEEE Communications Letters}, vol. 20, no. 12, pp. 2482-2485, Dec. 2016.
\bibitem{Jr10}\label{FD-WPCN-Mingjin}M. Gao, H. Chen, Y. Li, M. Shirvanimoghaddam and J. Shi, "Full-Duplex Wireless-Powered Communication with Antenna Pair Selection", accepted to appear in \textit{Proc. of~WCNC'15}, New Orleans, LA USA, 9-13 March, 2015.
\bibitem{Jr11}\label{FD-WPCN-Zihao}Z. Gao, H. Chen, Y. Li, B. Vucetic, "Wireless-Powered Communications with Two-Way Information Flow: Protocols and Throughput Regions", in \textit{Proc. of AusCTW'16}, Melbourne, Australia, 20-22, Jan. 2016.
\bibitem{Jr12}\label{FD-WPCN-Relay1}N. Zlatanov, D. W. K. Ng and R. Schober, "Capacity of the two-hop full-duplex relay channel with wireless power transfer from relay to battery-less source," in \textit{IEEE International Conference on Communications (ICC)}, 2016, pp. 1-7.
\bibitem{Jr13}\label{FD-WPCN-Relay2}Y. Zeng and R. Zhang, "Full-Duplex Wireless-Powered Relay With Self-Energy Recycling," in \textit{IEEE Wireless Communications Letters}, vol. 4, no. 2, pp. 201-204, April 2015.
\bibitem{Jr14}\label{FD-WPCN-Relay3}C. Zhong, H. A. Suraweera, G. Zheng, I. Krikidis and Z. Zhang, "Wireless Information and Power Transfer With Full Duplex Relaying," in \textit{IEEE Transactions on Communications}, vol. 62, no. 10, pp. 3447-3461, Oct. 2014.
\bibitem{Jr15}\label{FD-WPCN-Jamming}Y. Bi and H. Chen, "Accumulate and Jam: Towards Secure Communication via A Wireless-Powered Full-Duplex Jammer," in \textit{IEEE Journal of Selected Topics in Signal Processing}, vol.PP, no.99, pp.1-1, Aug. 2016.
\bibitem{Jr16}\label{EA1}Y. Gu, H. Chen, Y. Li and B. Vucetic, "A Discrete Time-Switching Protocol for Wireless-Powered Communications with Energy Accumulation," in \textit{2015 IEEE Global Communications Conference}, pp. 1-6, Dec 2015.
\bibitem{Jr17}\label{EA2}Y. Gu, H. Chen, Y. Li and B. Vucetic, "Distributed Multi-Relay Selection in Accumulate-then-Forward Energy Harvesting Relay Networks," \textit{ArXiv e-prints}, 2016.
\bibitem{bestH}\label{bestH}R. Knopp and P. A. Humblet, "Information capacity and power control in single-cell multiuser communications," in \textit{Proc. IEEE ICC}, pp. 331-335 vol.1, 1995.
\bibitem{PF1}\label{PF1}L. Yang and M. S. Alouini, "Performance Analysis of Multiuser Selection Diversity," in \textit{IEEE Transactions on Vehicular Technology}, vol. 55, no. 6, pp. 1848-1861, Nov. 2006.
\bibitem{hybridPF}\label{hybridPF}L. Yang, M. Kang and M. S. Alouini, "On the Capacity-Fairness Tradeoff in Multiuser Diversity Systems," in \textit{IEEE Transactions on Vehicular Technology}, vol. 56, no. 4, pp. 1901-1907, July 2007.
\bibitem{PF2}\label{PF2}Z. Hadzi-Velkov, I. Nikoloska, H. Chingoska and N. Zlatanov, "Proportional Fair Scheduling in Wireless Networks With RF Energy Harvesting and Processing Cost," in \textit{IEEE Communications Letters}, vol. 20, no. 10, pp. 2107-2110, Oct. 2016.
\bibitem{PF4}\label{PF4}Z. Hadzi-Velkov, I. Nikoloska, H. Chingoska and N. Zlatanov, "Opportunistic Scheduling in Wireless Powered Communication Networks," in \textit{IEEE Transactions on Wireless Communications}, vol.PP, no.99, pp.1-1.
\bibitem{Jr18}\label{Channel_training1}J. Xu and R. Zhang, "Energy Beamforming With One-Bit Feedback," in \textit{IEEE Transactions on Signal Processing}, vol. 62, no. 20, pp. 5370-5381, Oct.15, 2014.
\bibitem{Jr19}\label{Channel_training2}B. Hassibi and B. M. Hochwald, "How much training is needed in multiple-antenna wireless links?," in \textit{IEEE Transactions on Information Theory}, vol. 49, no. 4, pp. 951-963, April 2003.
\bibitem{Jr20}\label{iid1}I. Krikidis, T. Charalambous and J. S. Thompson, "Buffer-Aided Relay Selection for Cooperative Diversity Systems without Delay Constraints," in \textit{IEEE Transactions on Wireless Communications}, vol. 11, no. 5, pp. 1957-1967, May 2012.
\bibitem{Jr21}\label{Same_Eth}I. Krikidis, "Relay Selection in Wireless Powered Cooperative Networks With Energy Storage," in \textit{IEEE Journal on Selected Areas in Communications}, vol. 33, no. 12, pp. 2596-2610, Dec. 2015.
\bibitem{FD2}\label{FD2}V. Syrjala, M. Valkama, L. Anttila, T. Riihonen and D. Korpi, "Analysis of Oscillator Phase-Noise Effects on Self-Interference Cancellation in Full-Duplex OFDM Radio Transceivers," in \textit{IEEE Trans. on Wireless Commun.}, vol. 13, no. 6, pp. 2977-2990, June 2014.
\bibitem{FD3}\label{FD3}D. Bharadia, E. McMilin, S. Katti, "Full duplex radios", \textit{Proc. ACM SIGCOMM}, pp. 375-386, 2013.
\bibitem{FD5}\label{FD5}B. Radunovic et al., "Rethinking indoor wireless mesh design: Low power low frequency full-duplex", \textit{Proc. WIMESH}, pp. 1-6, 2010.
\bibitem{30}\label{math_tool}D. Zwillinger, \textit{Table of integrals, series, and products}. Elsevier, 2014.
\bibitem{markov_chain}\label{markov_chain}D.Freedman, \textit{Markov chains}. Holden-Day series in probability and statistics, 1971.
\bibitem{Fixed_point}\label{Fixed_point}J. M Ortega, W. C Rheinboldt, \textit{Iterative solution of nonlinear equations in several variable}. New York: Academic Press, 1970.
\bibitem{Rician_index}\label{Rician_index}J.G. Proakis, \textit{Digital Communications}, McGraw-Hill, 1995.
\bibitem{fairness_define}\label{fairness_define}R. Elliott, "A measure of fairness of service for scheduling algorithms in multiuser systems," in \textit{Proc. IEEE CCECE}, Winnipeg, MB, Canada, May, 2002, pp. 1583-1588.
\end{thebibliography}
\end{document}